\documentclass[pdflatex,sn-mathphys-num]{sn-jnl}

\usepackage{graphicx}%
\usepackage{multirow}%
\usepackage{amsmath,amssymb,amsfonts}%
\usepackage{amsthm}%
\usepackage{mathrsfs}%
\usepackage[title]{appendix}%
\usepackage{xcolor}%
\usepackage{textcomp}%
\usepackage{manyfoot}%
\usepackage{booktabs}%
\usepackage{algorithm}%
\usepackage{algorithmicx}%
\usepackage{algpseudocode}%
\usepackage{listings}%
\usepackage{svg}

\theoremstyle{thmstyleone}%
\theoremstyle{thmstyletwo}%

\theoremstyle{thmstylethree}%

\begin{document}

\title[Article Title]{JEDEL: Zero-Shot DNA-Encoded Library Design for Early-Stage Drug Discovery}


\author*[1]{\fnm{Zygimantas} \sur{Jocys}}\email{zj1g15@soton.ac.uk}
\author[1]{\fnm{Zhanxing} \sur{Zhu}}
\author[2]{\fnm{Henriette M.G.} \sur{Willems}}
\author[1]{\fnm{Katayoun} \sur{Farrahi}}

\affil*[1]{\orgdiv{School of Electronics and Computer Science}, \orgname{University of Southampton}, \orgaddress{\city{Southampton}, \postcode{SO17 1BJ}, \country{United Kingdom}}}
\affil[2]{\orgdiv{The ALBORADA Drug Discovery Institute}, \orgname{University of Cambridge}, \orgaddress{\street{Hills Road}, \city{Cambridge}, \postcode{CB2 0AH}, \country{United Kingdom}}}

\abstract{We present JEDEL, a framework for generating synthesis-ready DNA-encoded libraries (DELs) directly from three-dimensional pharmacophore representations of active ligands. JEDEL is the first model to map pharmacophore interaction patterns to actionable, scalable synthesis instructions, enabling the design of targeted libraries comprising potentially millions of molecules. Unlike existing generative approaches that produce virtual compounds requiring downstream synthesis planning, JEDEL operates within the space of purchasable building blocks and validated reactions, ensuring that every output is experimentally realizable by construction. JEDEL learns a predictive alignment between pharmacophore geometry and molecular structure and decodes this into combinatorial synthesis routes at scale. Across 18 protein targets, it generates focused libraries that outperform random and diversity-based baselines in predicted binding affinity, pharmacophore recovery, and sample efficiency, without target-specific retraining. JEDEL enables a shift from virtual molecule generation to experimentally deployable library design.}




\maketitle

\label{submission}

Generative molecular design has advanced rapidly, and yet a persistent gap remains between computational proposals and experimental validation. Most generative methods produce molecules that must be synthesized and tested one at a time, with no guarantee that proposed compounds are accessible through known chemistry.

Target-conditioned 3D models such as TargetDiff~\citep{guan20233dtargetdiff} and Pocket2Mol~\citep{Pocket2mol} generate molecules with favorable predicted binding poses. Benchmark evaluations show, however, that the majority of their outputs cannot be synthesized through realistic reaction schemes~\citep{Drugpose, luo2024projectingmoleculessynthesizablechemical}. Synthesis-aware approaches such as SynFlowNet~\citep{cretu2024synflownetmoleculedesignguaranteed} and SynNet~\citep{synnet} constrain outputs to feasible routes, though these methods depend on reward functions with limited accuracy and lack three-dimensional geometric awareness.

Experimental platforms face a complementary limitation. High-throughput screening operates on fixed physical libraries that cannot be refined in response to computational hypotheses. The result is a disconnect in which generative models explore chemical space freely, producing molecules that rarely reach the bench, while experimental platforms screen broadly without computational guidance.
\begin{figure}[t]
  \centering

  \makebox[\textwidth][l]{\sffamily\small\textbf{a}\hspace{4pt}\textit{Overview}}\par\vspace{1pt}
  \includegraphics[width=0.9\textwidth]{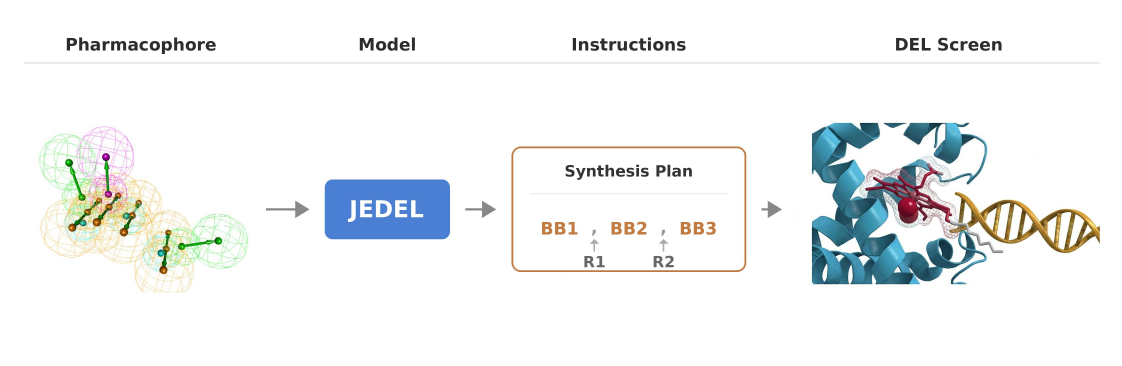}

  \vspace{2pt}
  {\color{black!30}\rule{0.9\textwidth}{0.25pt}}
  \vspace{2pt}

  \makebox[\textwidth][l]{\sffamily\small\textbf{b}\hspace{4pt}\textit{Architecture}}\par\vspace{1pt}
  \includegraphics[width=\textwidth]{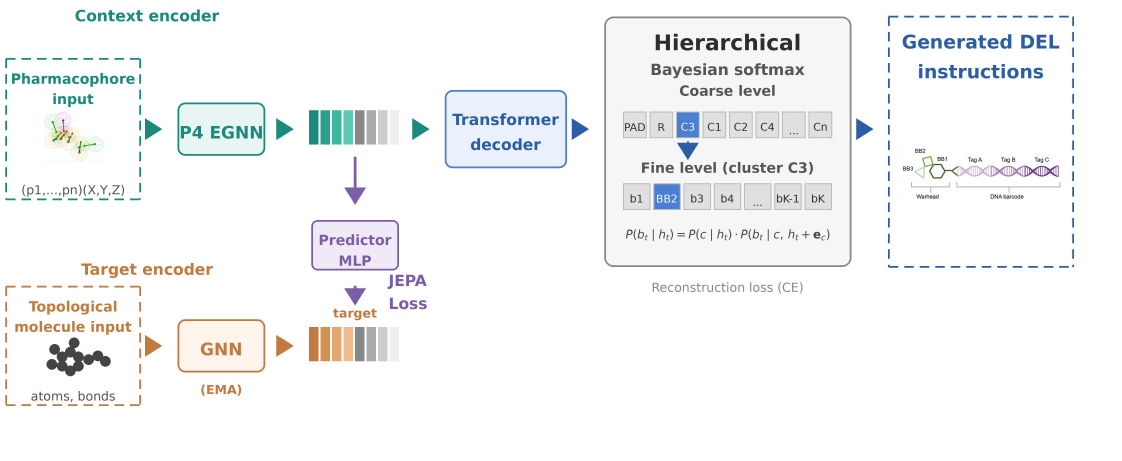}

  \caption{\textbf{JEDEL: a joint embedding predictive architecture for pharmacophore-guided DNA-encoded library design.}
  \textbf{a},~A 3D pharmacophore extracted from known active ligands is passed through JEDEL, which outputs a synthesis plan comprising purchasable building blocks (BB1--BB3) joined by compatible reactions (R1--R2), mapping directly to a DNA-encoded library compound ready for affinity screening.
  \textbf{b},~The context encoder (P4 EGNN) encodes 3D pharmacophore geometry while the target encoder (GNN, EMA-updated) encodes molecular topology; a predictor MLP aligns the two views via a JEPA loss. The context representation feeds a Transformer decoder whose hierarchical Bayesian softmax first selects a coarse cluster then a specific building block within it, scaling to vocabularies exceeding 200{,}000 purchasable reagents.}
  \label{fig:jedel}
\end{figure}

DNA-encoded libraries (DELs) offer a natural resolution to this disconnect. In the bead-based DEL paradigm, each molecule is assembled from purchasable building blocks through validated combinatorial reactions and linked to a DNA barcode on a single bead, enabling pooled screening and sequencing-based identification~\citep{cavett2015desps, price2017microfluidic, keller2024dual}. Unlike solution-phase split-and-pool libraries, which can reach billions of members but are constrained to affinity-based selection in aqueous conditions~\citep{clark2009design, brenner1992encoded}, bead-based DELs support activity-based and cellular screening modalities while maintaining unambiguous compound identity. This one-bead-one-compound format produces clean, well-defined experimental signals that are naturally suited to machine learning integration~\citep{mccloskey2020machine, iqbal2025evaluation}. Critically, the combinatorial construction from commercial reagents guarantees that every library member is synthesizable, provided the generative model operates within the space of valid building-block combinations and reaction schemes.
 
Yet no existing method generates focused DEL libraries conditioned on the binding requirements of a target of interest. The integration of machine learning with DELs has so far focused on post-screening analysis of split-and-pool data: denoising sequencing readouts, predicting binding from enrichment signals, and virtually screening beyond the physical library~\citep{McCloskey2020, Lim2022, Komar2020, Chen2024, Blevins2024}. ML-guided library \textit{design} for bead-based platforms, where the goal is to select which compounds to synthesize before any screening takes place, remains entirely unexplored.
 
A further challenge is that many therapeutically important targets lack reliable protein structures for structure-based design. Although AlphaFold~\citep{Jumper2021AlphaFold} has expanded structural coverage, inaccuracies in binding-site geometry and side-chain placement degrade design quality for flexible or membrane-associated targets. Ligand-based 3D pharmacophore representations, which encode the spatial arrangement of interaction features (hydrogen bond donors and acceptors, hydrophobic regions, aromatic centres) from known active compounds, provide a more robust and experimentally grounded design signal. Pharmacophore-conditioned generative models such as PGMG~\citep{zhu2022pgmg}, ShEPhERD~\citep{adams2025diffusing}, and PharmaDiff~\citep{alakhadar2025pharmadiff} have demonstrated the value of this conditioning, but none produce outputs compatible with any synthesis platform, let alone DEL construction.
 
We introduce JEDEL (Joint Embedding for DNA-Encoded Libraries), an architecture that translates 3D pharmacophore geometry directly into synthesis-ready DEL building-block sequences. JEDEL encodes pharmacophore point clouds using an E(n)-equivariant graph neural network \citep{satorras2021en} that respects rotational and translational symmetries, and separately encodes molecular topology through a message-passing graph neural network. A central design challenge is learning the mapping between these two representations: the correspondence between pharmacophore geometry and molecular structure is inherently one-to-many, since structurally diverse molecules can realise the same interaction pattern, and dissimilar scaffolds may share identical pharmacophore signatures~\citep{zhang2023activitycliffpredictiondataset}. Contrastive learning \cite{Chen2020SimCLR} is ill-suited to this setting because meaningful negatives cannot be defined in a space where structural similarity does not imply functional similarity. We therefore adopt a Joint Embedding Predictive Architecture (JEPA)~\citep{assran2023self}, which learns to predict latent topological representations from pharmacophore context through a predictive objective, naturally accommodating the one-to-many distribution while avoiding representation collapse through an asymmetric predictor-target design with an exponential moving average target encoder. A hierarchical decoder then translates these aligned embeddings into building-block selections from a vocabulary exceeding 200,000 purchasable reagents, guaranteeing that every output corresponds to an executable DEL synthesis protocol without post-hoc filtering.

\begin{figure}[h]
    \centering
    \includegraphics[width=\textwidth]{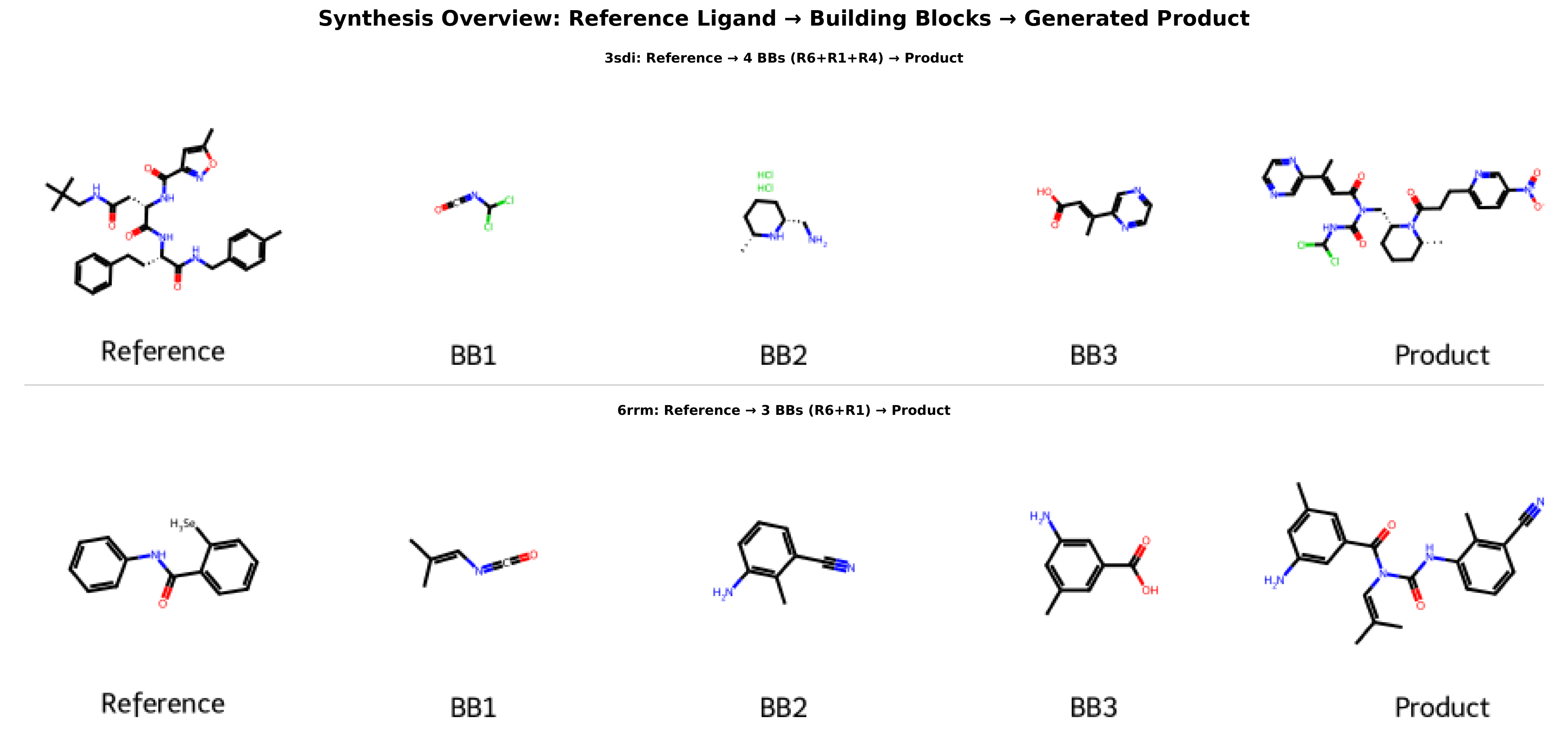}
    \caption{Representative molecules generated by JEDEL as DEL synthesis instructions, 
    shown alongside their reference ligand and constituent building blocks (BBs). 
    Each product is assembled via an enumerated reaction sequence (R1: amide coupling; 
    R4: C--C bond formation; R6: Ugi-type condensation) and is selected because its 
    four-point pharmacophore (p4) matches that of the reference. \textbf{Top (3SDI):} 
    The reference p4 is anchored by a central amide, a hydrophobic aromatic, and two 
    H-bond acceptors; the product recovers all four points via the Ugi condensation core 
    and the pyridine nitrogen, despite a fully distinct scaffold. \textbf{Bottom (6RRM):} 
    The reference benzamide presents a single H-bond donor/acceptor pair flanked by two 
    hydrophobic aromatics; the product preserves this geometry through the retained amide 
    and the tolyl/cyanobenzene substituents, using only three BBs. In both cases the 
    synthesis route is directly executable, making each generated molecule a concrete 
    DEL design proposal rather than a virtual structure.}
    \label{fig:synthesis_overview}
\end{figure}

We evaluate JEDEL across 18 diverse protein targets spanning proteases, kinases, GPCRs, nuclear receptors, and immune proteins. Without any target-specific retraining, JEDEL consistently outperforms both random DEL enumeration and a curated Enamine diversity library in predicted binding affinity, pharmacophore recovery, and sample efficiency, achieving 2.6-fold higher median hit rates than the diversity baseline. Every generated (See Figure \ref{fig:synthesis_overview}) molecule is immediately actionable in a bead-based DEL screening campaign. To our knowledge, JEDEL is the first framework to connect pharmacophore-driven molecular design with combinatorial library construction, closing the loop from computational hypothesis to experimental deployment.




\section{Results}

\subsection{Synthesis-ready generation with high chemical diversity}

\begin{figure}[ht]
    \centering
    \includegraphics[width=0.85\textwidth]{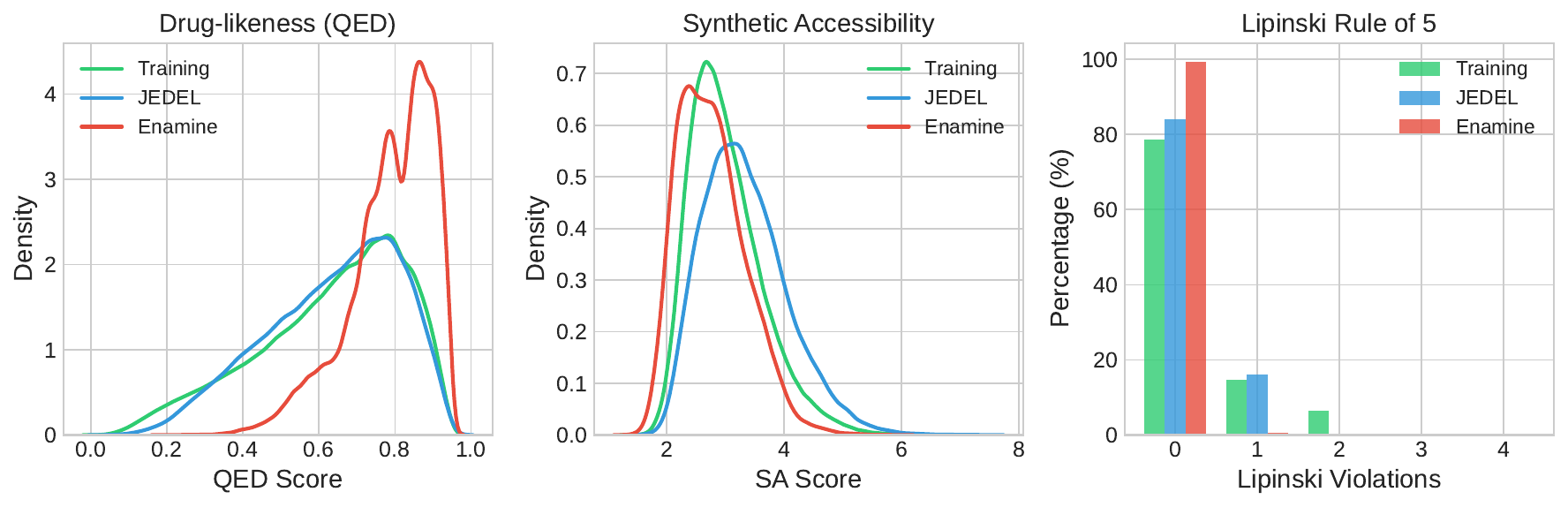}
    \par\vspace{0.5em}
    \includegraphics[width=0.85\textwidth]{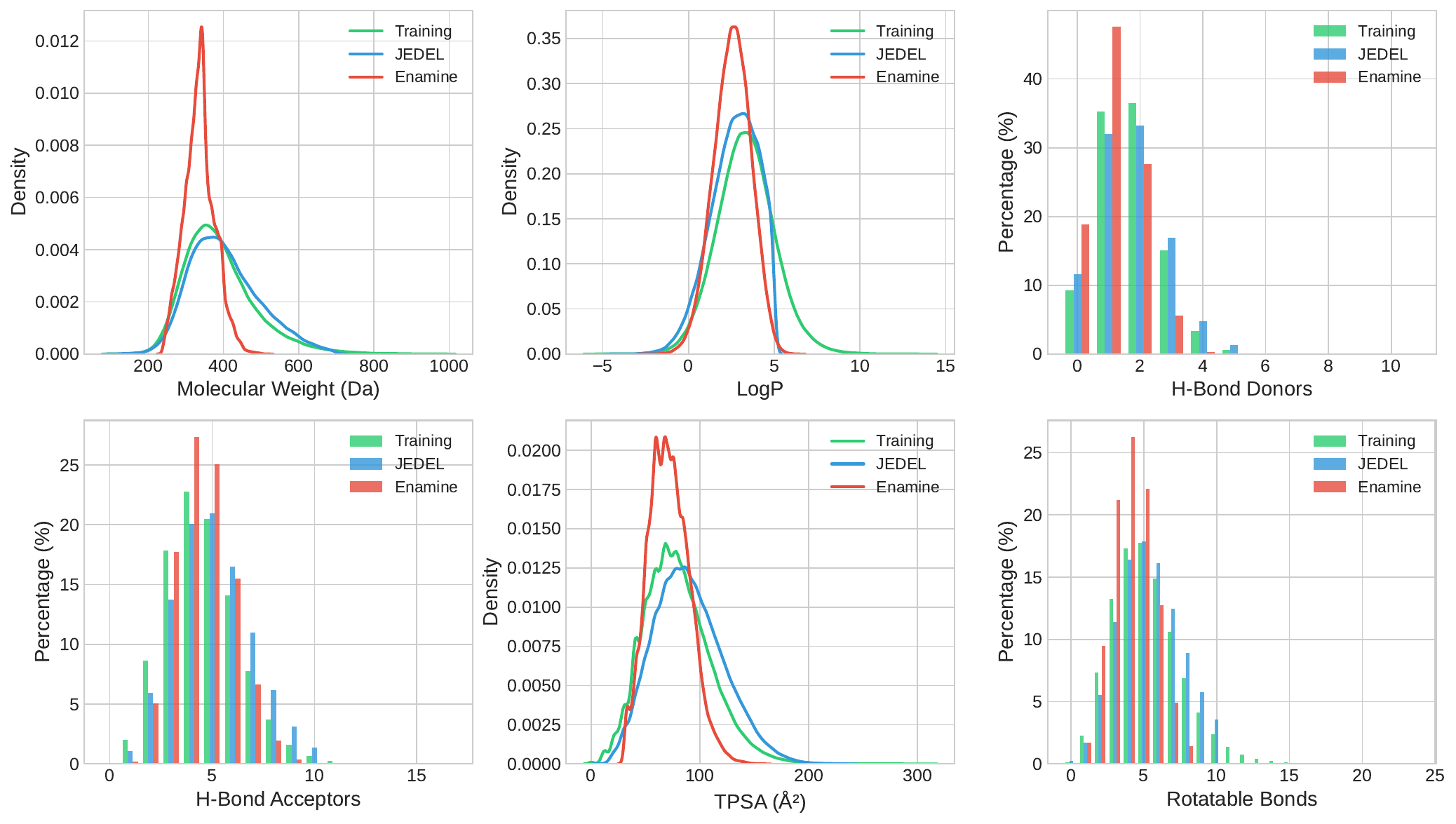}
    \caption{Drug-likeness and physicochemical property distributions of JEDEL-generated molecules compared with the training set and the Enamine diversity baseline. Top: QED score, synthetic accessibility (SA), and Lipinski Rule of 5 violations. Bottom: molecular weight, LogP, hydrogen bond donors and acceptors, topological polar surface area (TPSA), and rotatable bonds.}
    \label{fig:properties}
\end{figure}

All JEDEL-generated molecules are synthesisable by construction, a property guaranteed by the architecture rather than assessed post hoc. This constraint does not compromise chemical diversity: JEDEL achieves 100\% chemical validity, 99.3\% uniqueness, and 99.9\% novelty, with an internal diversity of 0.869 (Supplementary Table~2). Generated molecules exhibit drug-like profiles (mean QED $0.64 \pm 0.17$, 84\% Lipinski compliance, SA score $3.33 \pm 0.72$) and physicochemical properties consistent with the training distribution (mean MW $403 \pm 93$~Da, LogP $2.68 \pm 1.38$). Compared with the Enamine diversity baseline (MW $337 \pm 41$~Da), JEDEL produces somewhat larger molecules with greater hydrogen bonding capacity, consistent with pharmacophore-conditioned design targeting specific interaction patterns (Fig.~\ref{fig:properties}).

\subsection{Docking performance across 18 diverse protein targets}

Docking results are summarised in Table~\ref{tab:docking_p4}. Aggregated across all targets, JEDEL achieves the lowest mean docking energy ($-7.49$~kcal~mol$^{-1}$), outperforming both Enamine ($-7.39$~kcal~mol$^{-1}$) and Random sampling ($-6.28$~kcal~mol$^{-1}$), an average improvement of 1.2~kcal~mol$^{-1}$ over unguided generation.

At the per-target level, JEDEL attains the best mean docking score on the majority of evaluated proteins, including challenging systems such as 2w0j, 3dv1, 3mlb, 5khi, 5mmp, and 5nn0. The gains are most pronounced for targets with stronger reference binding energies, where JEDEL shifts the generated distribution towards lower predicted energies relative to both baselines. While Enamine performs competitively on several targets, its performance is less consistent across the full benchmark. In a small number of cases (1e34, 1snk), JEDEL generates molecules with docking scores comparable to or better than the co-crystallised ligand.

JEDEL identifies the single best-scoring molecule on 8 of 18 targets (44\%). The 5NN0 result is particularly notable: JEDEL's best molecule is 1.22~kcal~mol$^{-1}$ better than Enamine's best ($-10.03$ vs $-8.81$), suggesting that pharmacophore conditioning focuses sampling on chemically relevant regions of DEL space that brute-force enumeration misses.

\subsection{Pharmacophore recovery validates conditioning fidelity}

JEDEL achieves the strongest pharmacophore recovery across both metrics (Table~\ref{tab:docking_p4}). Aggregated across targets, JEDEL reaches mean P4 recovery of $89.3\% \pm 9.5$ (versus $88.1\% \pm 8.9$ for Enamine and $78.1\% \pm 9.6$ for Random) and the highest Jaccard similarity ($0.761 \pm 0.126$ versus $0.751 \pm 0.121$ and $0.694 \pm 0.083$, respectively).

The improvements are most pronounced on targets with complex pharmacophore signatures (Ref P4 = 5--6), including 1k9r, 1rp7, 5khi, and 5mmp, where coordinated recovery of multiple interaction types is required. Even on targets with low pharmacophore complexity (3dv1, 4ynk), JEDEL maintains performance comparable to or better than the baselines. The alignment between docking improvements and pharmacophore recovery across the benchmark indicates that JEDEL successfully integrates interaction-level constraints into molecular generation.

\begin{table*}[htbp]
\centering
\caption{Docking and pharmacophore (P4) recovery performance across 18 diverse protein targets. \textbf{Ref.} denotes the co-crystallised ligand binding energy (kcal/mol). \textbf{Ref P4} is the number of distinct pharmacophore feature types in the reference ligand. \textbf{Mean} is the mean predicted binding affinity. \textbf{Rec\%} is the mean P4 Recovery and \textbf{Jac} is the mean Jaccard similarity, each averaged over up to 100 generated molecules per method. Best value per target is shown in \textbf{bold}.}
\label{tab:docking_p4}
\resizebox{\textwidth}{!}{%
\begin{tabular}{lcc|ccc|ccc|ccc}
\toprule
& & & \multicolumn{3}{c|}{\textbf{Random}} & \multicolumn{3}{c|}{\textbf{Enamine}} & \multicolumn{3}{c}{\textbf{JEDEL}} \\
\textbf{Protein} & \textbf{Ref.} & \textbf{Ref P4} & Mean & Rec\% & Jac & Mean & Rec\% & Jac & Mean & Rec\% & Jac \\
\midrule
1e34 & $-5.93$ & 5 & $-5.44$ & 68.3 & 0.610 & $\mathbf{-6.65}$ & \textbf{76.5} & 0.649 & $-6.54$ & 76.3 & \textbf{0.664} \\
1k9r & $-6.18$ & 6 & $-4.63$ & 67.3 & 0.673 & $-5.68$ & 78.9 & 0.789 & $\mathbf{-5.88}$ & \textbf{83.3} & \textbf{0.833} \\
1rp7 & $-10.06$ & 6 & $-7.79$ & 67.3 & 0.673 & $-9.58$ & 78.9 & 0.789 & $\mathbf{-9.59}$ & \textbf{80.7} & \textbf{0.807} \\
1snk & $-6.54$ & 4 & $-5.54$ & 85.3 & 0.743 & $-6.72$ & 95.1 & 0.775 & $\mathbf{-7.17}$ & \textbf{97.0} & \textbf{0.783} \\
2lk1 & $-6.51$ & 4 & $-4.91$ & 63.7 & 0.465 & $\mathbf{-5.80}$ & \textbf{75.0} & \textbf{0.526} & $-5.54$ & 71.0 & 0.508 \\
2w0j & $-11.71$ & 5 & $-6.34$ & 80.7 & 0.807 & $-7.24$ & 94.3 & 0.940 & $\mathbf{-7.72}$ & \textbf{94.7} & \textbf{0.947} \\
2y1o & $-9.59$ & 5 & $-6.10$ & 68.3 & 0.610 & $-7.28$ & 76.5 & 0.649 & $\mathbf{-7.59}$ & \textbf{77.4} & \textbf{0.656} \\
3dv1 & $-9.19$ & 3 & $-6.48$ & 95.5 & \textbf{0.722} & $-7.73$ & 94.2 & 0.580 & $\mathbf{-8.65}$ & \textbf{97.4} & 0.588 \\
3kfc & $-9.38$ & 4 & $-7.44$ & 79.2 & 0.656 & $\mathbf{-8.75}$ & \textbf{97.2} & \textbf{0.809} & $-8.34$ & 94.5 & 0.769 \\
3mlb & $-13.03$ & 4 & $-6.80$ & 85.3 & 0.742 & $-7.54$ & 95.1 & 0.775 & $\mathbf{-7.74}$ & \textbf{98.0} & \textbf{0.786} \\
4i7f & $-8.73$ & 4 & $-7.49$ & 79.2 & 0.656 & $\mathbf{-8.45}$ & 97.2 & \textbf{0.809} & $-8.38$ & \textbf{99.7} & 0.804 \\
4x49 & $-4.58$ & 5 & $-5.34$ & 69.9 & 0.632 & $\mathbf{-5.97}$ & 75.2 & 0.629 & $-5.90$ & \textbf{75.5} & \textbf{0.641} \\
4ynk & $-14.70$ & 3 & $-7.37$ & \textbf{95.5} & \textbf{0.723} & $\mathbf{-8.93}$ & 94.2 & 0.580 & $-8.70$ & 96.4 & 0.593 \\
5jfu & $-11.18$ & 4 & $-6.35$ & 85.3 & 0.742 & $\mathbf{-8.11}$ & 95.1 & 0.775 & $-7.56$ & \textbf{95.2} & \textbf{0.781} \\
5khi & $-10.29$ & 6 & $-6.81$ & 67.3 & 0.673 & $-7.23$ & 78.9 & 0.789 & $\mathbf{-7.35}$ & \textbf{81.4} & \textbf{0.814} \\
5mmp & $-9.77$ & 5 & $-5.59$ & 80.7 & 0.807 & $-6.50$ & 94.3 & 0.940 & $\mathbf{-7.05}$ & \textbf{98.1} & \textbf{0.975} \\
5nbw & $-13.31$ & 4 & $-7.36$ & 85.3 & 0.742 & $\mathbf{-8.55}$ & 95.1 & 0.775 & $-8.19$ & \textbf{96.1} & \textbf{0.798} \\
5nn0 & $-7.10$ & 5 & $-5.20$ & 80.7 & 0.807 & $-6.38$ & 94.3 & 0.940 & $\mathbf{-6.95}$ & \textbf{94.7} & \textbf{0.942} \\
\midrule
\textit{Aggregate} & --- & --- & $-6.28$ & 78.1{\scriptsize$\pm$9.6} & 0.694{\scriptsize$\pm$0.083} & $-7.39$ & 88.1{\scriptsize$\pm$8.9} & 0.751{\scriptsize$\pm$0.121} & $\mathbf{-7.49}$ & \textbf{89.3}{\scriptsize$\pm$9.5} & \textbf{0.761}{\scriptsize$\pm$0.126} \\
\bottomrule
\end{tabular}%
}
\end{table*}

\subsection{Sample efficiency and hit rate analysis}

Hit rates, defined as the fraction of generated compounds with lower docking energy than the co-crystallised reference, are presented in Table~\ref{tab:sample-efficiency}. JEDEL outperforms or matches Enamine on the majority of targets with non-zero hits, with particularly large gains on 1k9r (50.0\% vs 20.0\%), 3dv1 (23.5\% vs 2.2\%), and 5nn0 (42.7\% vs 10.4\%). The median sample efficiency across targets with at least one hit is 14.0\% for JEDEL, compared with 5.4\% for Enamine and 0.1\% for Random, representing a 2.6-fold advantage over the curated diversity library.

Four targets (2w0j, 4ynk, 5mmp, 5nbw) yielded no compounds below the reference threshold for any method, reflecting intrinsic challenges in these binding sites or limitations of the docking scoring function.

\begin{table}[t]
\centering
\caption{Sample efficiency across target proteins. Values show the percentage of generated compounds with lower docking energy than the co-crystallised reference ligand. Bold indicates where JEDEL outperforms Enamine. Dashes indicate no method produced a compound below the reference threshold. Median computed over targets with $\geq$1 compound below threshold.}
\label{tab:sample-efficiency}
\small
\setlength{\tabcolsep}{10pt}
\begin{tabular}{lccc}
\toprule
\textbf{Protein} & \textbf{Random} & \textbf{Enamine} & \textbf{JEDEL} \\
\midrule
1e34 & 17.9\% & 89.5\% & 80.4\% \\
1k9r &  0.1\% & 20.0\% & \textbf{50.0\%} \\
1rp7 &  0.1\% & 31.8\% & \textbf{37.3\%} \\
1snk &  4.0\% & 59.1\% & \textbf{76.5\%} \\
2lk1 &  0.6\% & 10.9\% &  7.5\% \\
2w0j &  ---   &  ---   &  ---   \\
2y1o &  0.0\% &  0.1\% & \textbf{1.7\%} \\
3dv1 &  0.0\% &  2.2\% & \textbf{23.5\%} \\
3kfc &  2.0\% & 33.5\% & 20.1\% \\
3mlb &  0.0\% &  0.2\% &  0.0\% \\
4i7f & 12.0\% & 39.0\% & 35.8\% \\
4x49 & 79.4\% & 99.2\% & 97.3\% \\
4ynk &  ---   &  ---   &  ---   \\
5jfu &  0.0\% &  0.0\% &  0.0\% \\
5khi &  0.0\% &  0.2\% & \textbf{0.8\%} \\
5mmp &  ---   &  ---   &  ---   \\
5nbw &  ---   &  ---   &  ---   \\
5nn0 &  0.0\% & 10.4\% & \textbf{42.7\%} \\
\midrule
\textbf{Median} & \textbf{0.1\%} & \textbf{5.4\%} & \textbf{14.0\%} \\
\bottomrule
\end{tabular}
\end{table}

\subsection{Pharmacophore conditioning provides the greatest advantage on difficult targets}

Stratifying targets by difficulty (split at the median Enamine hit rate) reveals that JEDEL's advantage is largest where conventional screening fails. On difficult targets (Enamine hit rate $\leq 6.3\%$), JEDEL achieves a mean hit rate of 2.9\% compared with 0.3\% for Enamine, an order-of-magnitude improvement. On easier targets, JEDEL remains competitive (49.7\% vs 43.7\%). This pattern indicates that pharmacophore-guided generation provides the greatest value precisely when untargeted libraries lack sufficient chemical coverage.

\subsection{Ablation of the joint embedding objective}

The full model (JEPA plus reconstruction loss) achieves a reconstruction loss of 4.52 after 30 epochs, compared with 5.47 for an otherwise identical model trained without the JEPA objective. This 17\% improvement confirms that predictive alignment between pharmacophore and topological spaces provides an inductive bias that strengthens the decoder beyond what sequence-level supervision alone achieves.




\section{Discussion}

The most distinctive property of JEDEL is not a single quantitative improvement but a qualitative change in the design pipeline: every output is immediately actionable. Existing 3D generative models produce molecules that must be filtered through retrosynthetic analysis, vendor search, and route optimisation, discarding up to 90\% of candidates in the process\cite{Drugpose}. JEDEL eliminates this gap entirely. Each output corresponds to purchasable building blocks and validated DEL reaction schemes, making the transition from computation to experiment direct and lossless. This synthesis-by-construction guarantee is architectural rather than heuristic, and it applies uniformly across all generated molecules regardless of target or complexity.

This guarantee has immediate practical implications for bead-based DEL campaigns, where each molecule carries real synthesis cost and bead occupancy. JEDEL's 2.6-fold sample efficiency advantage over the Enamine diversity library means that smaller, more focused libraries can be constructed with a higher expected fraction of active compounds. Rather than synthesising large unfocused libraries and relying on statistical coverage, JEDEL enables hypothesis-driven library construction at the $10^3$--$10^4$ scale typical of bead-based platforms, where every compound is selected to satisfy specific pharmacophoric requirements.

JEDEL achieves this performance across 18 structurally diverse protein families without any target-specific retraining. This zero-shot capability is enabled by conditioning on ligand-derived pharmacophore features rather than protein structures, making the approach robust to targets with limited or inaccurate structural information. The consistency of improvements across proteases, kinases, GPCRs, nuclear receptors, and immune targets indicates that pharmacophore conditioning provides a generalisable design signal rather than one tuned to particular binding site geometries.

The pattern of where JEDEL helps most is informative about the nature of pharmacophore conditioning itself. The strongest advantages appear on targets with well-defined directional interactions: on 3DV1 (aspartic protease), the 0.92~kcal~mol$^{-1}$ improvement over Enamine reflects successful capture of the catalytic dyad hydrogen bonding pattern, while large gains on 5MMP and 5NN0 indicate that the model identifies specific interaction requirements that diversity screening misses. Conversely, where JEDEL underperforms Enamine, as with 2LK1 and 3KFC, the pharmacophore signal is inherently less discriminative. In 2LK1, the binding site is a shallow, solvent-exposed surface dominated by aromatic stacking against a zinc-coordinated motif, which standard pharmacophore features encode poorly. In 3KFC, the pocket is deep but overwhelmingly hydrophobic and flexible enough to accommodate different scaffolds in opposing orientations, meaning the pharmacophore template degenerates into a generic shape constraint. Both targets share only four distinct pharmacophore feature types, among the lowest in the benchmark, reducing the information content of conditioning to the point where Enamine's combinatorial coverage compensates through sheer scale. JEDEL's pharmacophore guidance adds the most value in a regime of intermediate binding site complexity: specific enough to benefit from guided generation, but not so nonspecific that brute-force enumeration covers the space equally well.

The hierarchical decoder contributes to JEDEL's practical scalability. By factorising predictions into cluster-level and building-block-level stages, it reduces output dimensionality from $\mathcal{O}(|V|)$ to $\mathcal{O}(C + K)$, a 100$\times$ reduction that enables generation of $10^6$ molecules on a single GPU. This matches the scale of typical DEL selections while providing pharmacophore-guided focusing unavailable in random enumeration. Compared with traditional high-throughput screening, which operates on fixed physical libraries with no feedback loop, JEDEL enables iterative computational refinement before any synthesis takes place.

Several limitations should be noted. Our evaluation relies on docking scores as a proxy for binding affinity, which may not capture entropic contributions, water-mediated contacts, or protein flexibility. The current implementation supports seven DEL-compatible reactions; expanding this coverage would broaden the accessible chemical space but requires additional validation. For targets with very high reference ligand affinities ($>$13~kcal~mol$^{-1}$), neither JEDEL nor baseline methods consistently produce superior molecules, suggesting that current building block diversity may be insufficient for highly optimised binding sites. Most critically, experimental validation through bead-based DEL synthesis and affinity screening is needed to confirm that computational predictions translate to measured binding activity. Future directions include active learning loops that integrate screening feedback into iterative library refinement, multi-pharmacophore conditioning for polypharmacology applications, and extension to split-and-pool DEL platforms.

In summary, JEDEL establishes the first practical connection between pharmacophore-driven molecular design and combinatorial library construction. By embedding both structural and synthetic constraints directly into the generative model, it produces focused, synthesis-ready libraries that consistently outperform conventional approaches in predicted binding quality, pharmacophore fidelity, and sample efficiency across diverse protein targets. Every generated molecule is immediately executable in a bead-based DEL screening campaign, providing a direct pathway from computational hypothesis to experimental deployment.


\section{Methods}

\subsection{Problem formulation}

We formulate focused DEL library generation as a conditional sequence generation problem. Given a reference ligand $m_{\text{ref}}$ with known activity, we compute its pharmacophore $\mathbf{p} = \texttt{Pharm}(m_{\text{ref}})$ encoding the 3D spatial arrangement of interaction features. A learned encoder maps this to a latent representation $\mathbf{z} = g_\phi(\mathbf{p})$, and a conditional generator $p_\theta(\mathbf{s} \mid \mathbf{z})$ produces synthesis routes such that the resulting library:
\begin{equation}
    \mathcal{L} = \left\{\texttt{Assemble}(\mathbf{s}_i) \;:\; \mathbf{s}_i \sim p_\theta(\cdot \mid g_\phi(\mathbf{p})),\; i = 1, \ldots, n\right\}
\end{equation}
satisfies \textit{pharmacophoric relevance} (generated molecules share the interaction features of $m_{\text{ref}}$) and \textit{structural diversity} (the library covers a broad range of scaffolds around that shared pharmacophore). The model is trained on pharmacophore--molecule pairs without protein target information and generalises to unseen targets at inference.

In the bead-based DEL setting, each molecule $m$ is encoded on a single bead via its synthesis route $\mathbf{s} = (b_1, b_2, \ldots, b_T)$, where each $b_t \in \mathcal{B}_t$ is a building block drawn from a position-specific catalogue and $T$ is the number of synthesis steps. The final molecule is obtained by executing the route: $m = \texttt{Assemble}(\mathbf{s})$.

\subsection{Architecture}

JEDEL comprises a pharmacophore encoder, a topological encoder, a predictor network, and a hierarchical decoder (Fig.~\ref{fig:jedel}). Three requirements motivate the design: (i) reasoning over 3D pharmacophore geometry while respecting rotational and translational symmetries, (ii) capturing the many-to-many correspondence between pharmacophore patterns and molecular topology, and (iii) constraining every output to a valid synthesis protocol over a vocabulary exceeding $2 \times 10^{5}$ building blocks.

\subsubsection{Pharmacophore encoder}

The pharmacophore encoder processes 3D pharmacophore points using an E($n$)-equivariant graph neural network (EGNN)\citep{satorras2021en}. We represent pharmacophores across seven types: hydrophobic, aromatic, hydrogen bond acceptor, hydrogen bond donor, positive ionizable, negative ionizable, and unknown. Each point is initialised with a learnable type embedding $\mathbf{h}_i^{(0)} = \text{Embed}(t_i) \in \mathbb{R}^{256}$. The EGNN propagates information through message passing:
\begin{align}
\mathbf{m}_{ij} &= \phi_e\left(\mathbf{h}_i^{(l)}, \mathbf{h}_j^{(l)}, \|\mathbf{x}_i - \mathbf{x}_j\|^2, a_{ij}\right) \\
\mathbf{h}_i^{(l+1)} &= \mathbf{h}_i^{(l)} + \phi_h\left(\mathbf{h}_i^{(l)}, \sum_{j \neq i} \tilde{a}_{ij} \mathbf{m}_{ij}\right)
\end{align}
where $\phi_e$ and $\phi_h$ are MLPs with SiLU activation and layer normalisation, and $\tilde{a}_{ij}$ are learned attention weights. The P4-EGNN uses $L_p = 7$ layers with hidden dimension 256. Node representations are aggregated into a global embedding $\mathbf{z}_p$ via attention-weighted pooling:
\begin{equation}
\mathbf{z}_p = \sum_{i=1}^{N_p} \alpha_i \mathbf{h}_i^{(L_p)}, \quad \alpha_i = \frac{\exp(\mathbf{w}^\top \mathbf{h}_i^{(L_p)})}{\sum_j \exp(\mathbf{w}^\top \mathbf{h}_j^{(L_p)})}
\end{equation}

\subsubsection{Topological encoder}

The topological encoder processes the 2D molecular graph using a message-passing GNN with $L_t = 7$ layers. Given atom features $\mathbf{a} \in \mathbb{Z}^{N_a}$ (11 atom types) and adjacency matrix $\mathbf{A}$:
\begin{align}
\mathbf{m}_{ij}^{(l)} &= \psi_m\left([\mathbf{h}_i^{(l)} \| \mathbf{h}_j^{(l)}]\right) \cdot \mathbf{A}_{ij} \\
\mathbf{h}_i^{(l+1)} &= \psi_u\left([\mathbf{h}_i^{(l)} \| \bar{\mathbf{m}}_i^{(l)}]\right) + \mathbf{h}_i^{(l)}
\end{align}
where $\bar{\mathbf{m}}_i^{(l)}$ is the mean-aggregated neighbourhood message and $\psi_m$, $\psi_u$ are two-layer MLPs. The encoder produces $\mathbf{z}_t \in \mathbb{R}^{256}$ via attention-weighted pooling.

\subsubsection{Joint embedding predictive architecture}

The mapping between pharmacophore geometry and molecular topology is inherently many-to-many: structurally diverse molecules can realise the same interaction pattern, and dissimilar scaffolds may share identical pharmacophore signatures\cite{zhang2023activitycliffpredictiondataset}. Contrastive objectives are ill-suited because meaningful negatives cannot be defined in molecular space. We therefore adopt a predictive objective inspired by JEPA\citep{assran2023self}.

Let $\mathbf{z}_c = [\mathbf{z}_{\text{3D}} \| \mathbf{z}_p]$ denote the context embedding formed by concatenating the 3D structure embedding (from a separate EGNN with $L_c = 4$ layers) and the pharmacophore embedding. A predictor $p_\psi$ (2-layer MLP, hidden dimension 256, GELU, dropout) maps the context to the target space:
\begin{equation}
\hat{\mathbf{z}}_t = p_\psi(\mathbf{z}_c)
\end{equation}
The objective minimises cosine distance between predicted and target embeddings:
\begin{equation}
\mathcal{L}_{\text{JEPA}} = 1 - \frac{\langle \hat{\mathbf{z}}_t, \text{sg}[\mathbf{z}_t^{\text{EMA}}] \rangle}{\|\hat{\mathbf{z}}_t\| \cdot \|\mathbf{z}_t^{\text{EMA}}\|}
\end{equation}
where $\text{sg}[\cdot]$ is the stop-gradient operator and $\mathbf{z}_t^{\text{EMA}}$ is computed by an exponential moving average target encoder ($\tau = 0.996$).

\subsubsection{Hierarchical decoder}

Computing a softmax over 217,067 building blocks at each decoding step is prohibitively expensive. We organise building blocks into 215 clusters based on reaction role (nucleophile/electrophile), primary compatible reaction (R1--R7), and pharmacophore profile (feature counts binned into four categories across six types). Clusters exceeding 1,024 members are split; those below 10 are merged by reaction type.

At each position $t$, the decoder predicts at two levels: a coarse distribution over token types $c_t \in \{\text{special, reaction, cluster}_1, \ldots, \text{cluster}_C\}$, and, for building block clusters, a fine distribution over individual blocks within the selected cluster:
\begin{equation}
P(b_t | \mathbf{h}_t) = P(c_t = c | \mathbf{h}_t) \cdot P(b_t | c_t = c, \mathbf{h}_t + \mathbf{e}_c)
\end{equation}
where $\mathbf{e}_c \in \mathbb{R}^{256}$ is a learnable cluster embedding. The decoder is a 4-layer Transformer with causal self-attention, cross-attention to the encoder output, 8 attention heads, feed-forward dimension 1,024, GELU activation, and dropout 0.1. This reduces output dimensionality from $O(|V|)$ to $O(C + K)$, a 100$\times$ reduction.

\subsection{Training}

The total loss combines joint embedding and hierarchical reconstruction objectives:
\begin{equation}
\mathcal{L} = \lambda_{\text{JEPA}} \mathcal{L}_{\text{JEPA}} + \lambda_{\text{rec}} \left( \mathcal{L}_{\text{coarse}} + \mathcal{L}_{\text{fine}} \right)
\end{equation}
with $\lambda_{\text{JEPA}} = \lambda_{\text{rec}} = 1.0$. The fine-level loss is computed only at building block positions. Training uses AdamW ($\text{lr} = 10^{-4}$), batch size 32, cosine annealing, for 50 epochs on a single NVIDIA V100 GPU. Model selection is based on validation loss.

At inference, pharmacophore features are extracted from a co-crystallised ligand, encoded via the P4-EGNN, projected to decoder space ($\mathbf{h}_0 = W_{\text{proj}} \mathbf{z}_c$), and the synthesis route is sampled autoregressively with temperature-controlled hierarchical decoding. Token sequences are converted to SMILES via RDKit reaction execution.

\subsection{Data}

\subsubsection{Building block library and reaction chemistry}

The building block library is constructed from 279,781 commercially available Enamine compounds designed for DEL synthesis. DEL-compatible functional groups are identified via SMARTS pattern matching (primary/secondary amines, carboxylic acids, boronic acids/esters, aldehydes, aryl halides, sulfonyl chlorides, isocyanates), yielding 217,067 retained blocks (80.8\%; MW 54--500~Da, mean 223~Da).

Seven DEL-compatible reactions are implemented: amide bond formation (R1, R2; 175,316 blocks), reductive amination (R3; 111,184), Suzuki-Miyaura coupling (R4; 80,189), click chemistry (R5; 4,617), sulfonamide formation (R6; 105,662), and urea formation (R7; 101,540). Each is implemented as a SMARTS pattern in RDKit with automatic reactant ordering.

\subsubsection{Training data}

Synthesis routes are enumerated by iteratively combining building blocks through compatible reactions (2--5 blocks per route, 1--4 reaction steps), validated by SMILES parsing and molecular weight constraints (maximum 1,000~Da). We sample 8 million training and 100,000 validation routes, balanced across complexity levels. For each molecule, a single low-energy 3D conformer is generated using ETKDG\citep{riniker2015better} with MMFF94 optimisation. Pharmacophore features are extracted via SMARTS-based pattern matching across six types, with aromatic ring atoms aggregated to centroids and nearby hydrophobic features merged within 1.5~\AA{}.

\subsection{Evaluation}

\subsubsection{Protein targets}

We evaluate on 18 targets from PDBbind\citep{wang2004pdbbind}, selected by one-per-family stratified sampling with MaxMin diversity optimisation. The benchmark spans proteases, kinases, GPCRs, nuclear receptors, transporters, transferases, and DNA-binding proteins (Supplementary Table~1). Mean pairwise ligand Tanimoto distance is 0.903.

\subsubsection{Docking protocol}

Virtual screening uses Smina\citep{koes2013lessons} with exhaustiveness 8, a single binding mode per ligand, and 120-second timeout. Ligands are prepared via ETKDG, MMFF optimisation, and conversion to PDBQT format. The search box is centred on the co-crystallised ligand with 8~\AA{} padding.

\subsubsection{Baselines}

All methods are compared under a matched budget of 100,000 compounds per target. Random DEL screening uniformly samples from the DEL chemical space. Diversity library screening uses the Enamine 100k diversity set with identical docking protocols.

\subsubsection{Metrics}

A molecule is classified as a hit if its docking score falls below the reference compound energy. We report hit rate, enrichment factor, and best docking scores. Generation quality is assessed via validity, uniqueness, novelty, and internal diversity. Drug-likeness is characterised by QED, SA score, and Lipinski compliance.

Pharmacophore fidelity is measured by P4 Recovery (fraction of reference pharmacophore types present in the generated molecule) and Jaccard Similarity (penalising extraneous types):
\begin{equation}
\text{Recovery}(m) = \frac{|\mathcal{F}_{\text{ref}} \cap \mathcal{F}_{m}|}{|\mathcal{F}_{\text{ref}}|}, \quad J(m) = \frac{|\mathcal{F}_{\text{ref}} \cap \mathcal{F}_{m}|}{|\mathcal{F}_{\text{ref}} \cup \mathcal{F}_{m}|}
\end{equation}
computed over the top 100 molecules per target per method.

\bibliography{sn-bibliography}

\end{document}